\documentclass[aps,prb,twocolumn,superscriptaddress,showkeys]{revtex4-1}

\usepackage{color}

\usepackage{graphicx}
\usepackage{dcolumn}
\usepackage{bm}

\usepackage[english]{babel}
\usepackage[latin1]{inputenc}
\usepackage[bottom=2cm,top=2cm, left=1.5cm, right=1.5cm]{geometry}

\usepackage[unicode=true, bookmarks=true, bookmarksnumbered=false, bookmarksopen=false, breaklinks=false, pdfborder={0 0 1}, backref=false, colorlinks=false]{hyperref}
\usepackage[sort&compress]{natbib}

\begin{document}

\title{Nanoscale nonlinear effects in Erbium-implanted Yttrium Orthosilicate}

\author{Nadezhda Kukharchyk}
\affiliation{Angewandte Festk\"{o}rperphysik, Ruhr-Universit\"{a}t Bochum, D-44780 Bochum, Germany}
\email{nadezhda.kukharchyk@ruhr-uni-bochum.de}

\author{Stepan Shvarkov}
\affiliation{Optoelektronische Materialien und Bauelemente, Universit\"{a}t Paderborn, D-33098 Padeborn, Germany}

\author{Sebastian Probst}
\affiliation{Quantronics group, Service de Physique de l'Etat Condense, DSM/IRAMIS/SPEC, CNRS UMR 3680, CEA-Saclay, 91191 Gif-sur-Yvette cedex, France}

\author{Kangwei Xia}
\affiliation{3. Physikalisches Institut, Universit\"{a}t Stuttgart, D-70569 Stuttgart, Germany}

\author{Hans-Werner Becker}
\affiliation{RUBION, Ruhr-Universit\"{a}t Bochum, D-44780 Bochum, Germany}

\author{Shovon Pal}
\affiliation{Angewandte Festk\"{o}rperphysik, Ruhr-Universit\"{a}t Bochum, D-44780 Bochum, Germany}
\affiliation{AG THz Spectroscopie und Technologie, Ruhr-Universit\"{a}t Bochum, D-44780 Bochum, Germany}
	
\author{Sergej Markmann}
\affiliation{AG THz Spectroscopie und Technologie, Ruhr-Universit\"{a}t Bochum, D-44780 Bochum, Germany}

\author{Roman Kolesov}
\affiliation{3. Physikalisches Institut, Universit\"{a}t Stuttgart, D-70569 Stuttgart, Germany}
\author{Petr Siyushev}
\affiliation{3. Physikalisches Institut, Universit\"{a}t Stuttgart, D-70569 Stuttgart, Germany}
\author{J\"{o}rg Wrachtrup}
\affiliation{3. Physikalisches Institut, Universit\"{a}t Stuttgart, D-70569 Stuttgart, Germany}

\author{Arne Ludwig}
\affiliation{Angewandte Festk\"{o}rperphysik, Ruhr-Universit\"{a}t Bochum, D-44780 Bochum, Germany}

\author{Alexey V. Ustinov}
\affiliation{Physikalisches Institut, Karlsruhe Institute of Technology, D-76128 Karlsruhe, Germany}
\author{Andreas D. Wieck}
\affiliation{Angewandte Festk\"{o}rperphysik, Ruhr-Universit\"{a}t Bochum, D-44780 Bochum, Germany}
\author{Pavel Bushev}
\affiliation{Experimentalphysik, Universit\"{a}t des Saarlandes, D-66123 Saarbr\"{u}cken, Germany}

\date{\today}

\begin{abstract}
	Doping of substrates at desired locations is a key technology for spin-based quantum memory devices. Focused ion beam implantation is well-suited for this task due to its high spacial resolution.
	In this work, we investigate ion-beam implanted erbium ensembles in Yttrium Orthosilicate crystals by means of confocal photoluminescence spectroscopy. 
	The sample temperature and the post-implantation annealing step strongly reverberate in the properties of the implanted ions. 
	We find that hot implantation leads to a higher activation rate of the ions. At high enough fluences, the relation between the fluence and final concentration of ions becomes non-linear. 
	Two models are developed explaining the observed behaviour.  
\end{abstract}

\maketitle

\section{Introduction}
The development of systems and devices suitable for quantum communication protocols has largely advanced during the last two decades. These systems are optimised for a single application such as data processing, data transmission or data storage. In order to develop a feasible quantum-information infrastructure, it is necessary to combine the best properties of those individual quantum systems within a so-called hybrid system \cite{georgescu2014,kurizki2015}. 
An example of a promising hybrid system are nuclear and electron spins coupled to a superconducting resonators and qubits \cite{imamoglu2009, kubo2010,kubo2011}.
A great leap towards the realization of such an interface had been made in circuit QED experiments with a variety of spin-ensembles \cite{ranjan2013,schuster2010,kubo2010,amsuss2011,rabl2006,wesenberg2009,probst2013,probst2014}. Spin-ensembles can be prepared locally on the substrates, thus allowing an arrangement of several bit-units with customised properties on one chip.
This enables a simple and available technological process for fabrication of quantum circuits.

We would like to emphasise the ensembles of rare-earth ions (RE) due to their unique properties. The optically active 4f-electrons of the RE elements are semi-shielded from the external fields by the 5d and 6s orbitals, which results in long optical and microwave coherence times.
Kramers RE ions possess large magnetic moments, which can be used for the manipulation of their spin states with microwaves. 
The realization of such spin ensembles by focused ion beam (FIB) implantation technique has many advantages, such as freedom in positioning of different ensembles in one crystal, variability of their concentrations and design of their arrangement in a single mask-less process \cite{kukharchyk2014}. 

One of the crucial requirements for the host substrates is a low local symmetry of the surrounding oxygen ions, which stabilises the RE ions in the lattice \cite{ishii2003,isshiki2003}. This property is inherent in Y$_2$SiO$_5$, Y$_3$Al$_5$O$_{12}$, YAlO$_3$ and YVO$_4$ crystals. 
In this work, we present in-depth analysis of the FIB-implanted erbium ensembles in Y$_2$SiO$_5$ (YSO) crystals.
This crystal is known to have the longest measured optical coherence time of about 4\,ms inside the telecom C-band at a wavelength of about 1.5\,$\mu$m \cite{Bottger2009, Bottger2003}.
However, formation of defects and dislocations during the implantation process aggravates the optical properties of the ions. Therefore, a deep understanding of ion-matter interactions and activation rates in a particular substrate is indispensable.
The first results on the activation rate of the implanted Erbium ensembles have already been reported in \cite{kukharchyk2014}. 

In this work, we demonstrate a more fundamental analysis of the focused ion beam implanted Erbium ensembles in the Y$_2$SiO$_5$ (YSO) crystals in dependence on the preparation process. 
The implantation of YSO crystals is performed in a wide range of fluences at different temperatures of the substrate. 
Optical properties of the implanted ions are analysed in dependence on fluence, ion concentration and ion activation by means of confocal photoluminescence.
We compare the results of the photolumisence measurements to the results of microwave spectroscopy of the implanted samples \cite{probst2014}.

\begin{figure*}[t]
	\includegraphics[width=\textwidth]{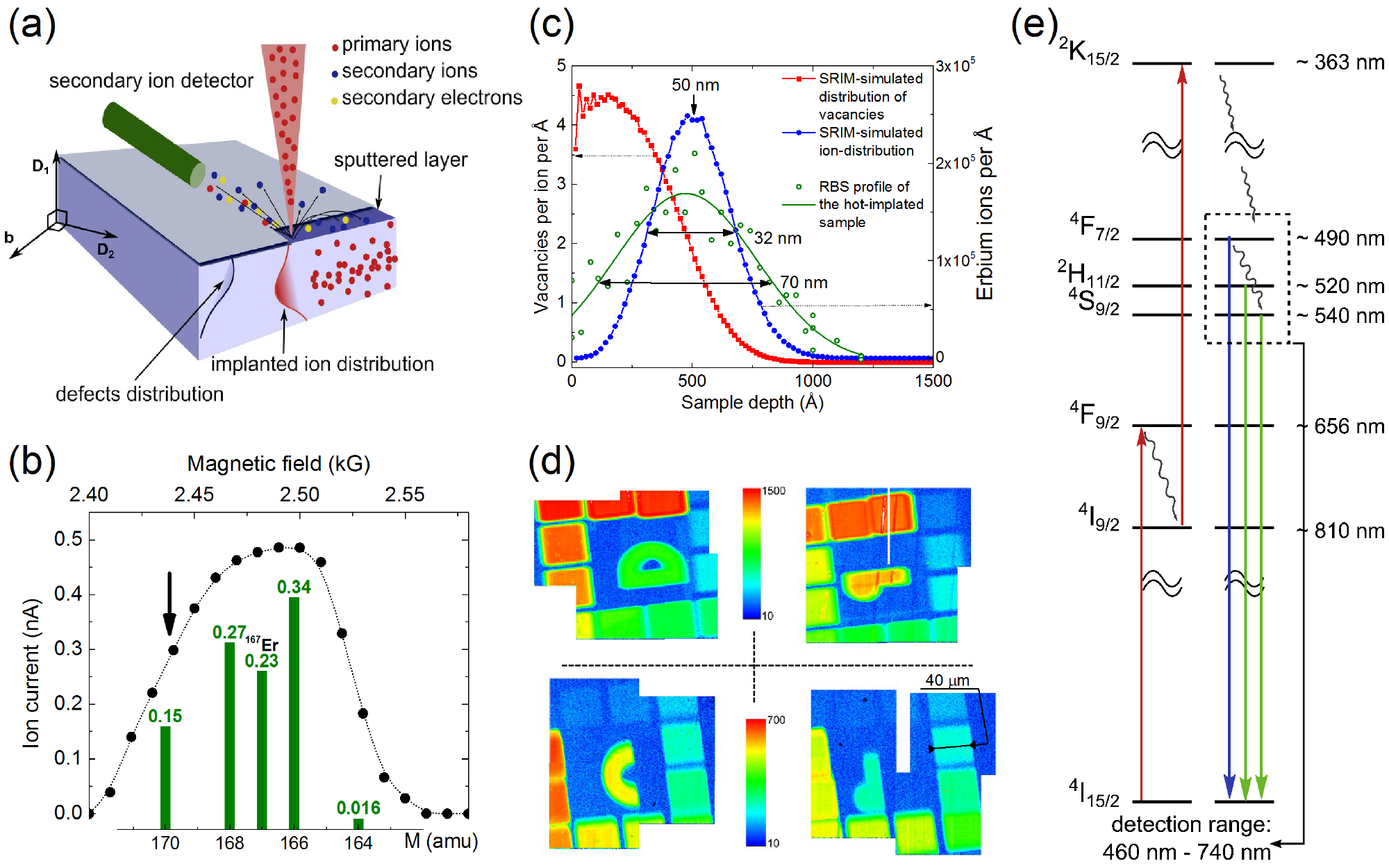}
	\caption[Sample preparation and measurement methods]{\textbf{(a)} Schematic of the implantation process. The orientation of the optical symmetry axes is shown accordingly to the sample fabrication and direction of the ion-beam. The ion channelling direction coincides with the D2-axis and is perpendicular to the implantation direction. \textbf{(b)} Mass-spectrum of the Er$^{++}$ peak. The distribution of Erbium isotopes is given by bars according to their natural abundance. The arrow points to the selected position for the implantation. \textbf{(c)} SRIM-simulated data for the ion-distribution (blue) and the vacancy-distribution (red) and RBS measured ion-distribution (green). The RBS result corresponds to the fluence of $2\times10^{15}$\,cm$^{-2}$, below which the concentration is under the detection limit. Depth of the ion-maximum is 50\,nm for both SRIM and RBS results. \textbf{(d)} Scanned luminescence pictures of the implanted patterns, intensity in Counts. Letters a, b, c, and d in the middle of patterns correspond to the implantation process as follows: a - implanted at 600\,K and annealed; b - implanted at 300\,K and annealed; c - implanted at 600\,K and not annealed; d - implanted at 300\,K and not annealed. \textbf{(e)} Level-diagram of an Er$^{3+}$ ion. Excitation is schematically shown with the red arrows.}
	\label{fig1}
\end{figure*}

\section{Experimental techniques}
\textbf{Focused ion beam implantation}

Samples are implanted in a focused ion beam (FIB) machine EIKO-100 with an accelerating voltage of 100\,kV under high-vacuum conditions of $10^{-7}$\,mBar. Schematic of the implantation process is demonstrated in Fig.\,\ref{fig1}\,(a).
In the current work, the FIB operates with a Liquid Metal Alloy Ion Source (LMAIS) Au$_{78.4}$Si$_{11.6}$Er$_{10}$ developed by Melnikov et al. \cite{melnikov2002}. Ions are separated by a built-in Wien mass-filter with a resolution of 2\,a.m.u. For the Er$^{++}$ peak, Fig.\,\ref{fig1}\,(b), it is possible to make a selection of the isotopic mixture with the lowest content of $^{167}$Er (to avoid the spectral broadening from the hyperfine structure). The distribution of the Erbium isotopes relative to the mass spectrum peak is shown in Fig.\,\ref{fig1}\,(b). In this work, the 90\%-pure $^{170}$Er$^{++}$ isotope is implanted with a Wien-filter setting as shown with an arrow in Fig.\,\ref{fig1}\,(b).

For YSO crystals, the sputter rate is low due to the high binding energies of the ions \cite{ulrich2003,rubio2014}. The distributions of ions and induced vacancies are simulated with the SRIM software \cite{Ziegler2010}, Fig.\,\ref{fig1}\,(c). As the channelling direction coincides with the D2-axis and is perpendicular to the direction of the ion-beam, Fig.\,\ref{fig1}\,(a), the SRIM simulation is applicable for the ion-distribution. From the ion- and defect-profiles, we estimate final ion-distribution to be 50\,nm deep for the ion maximum and 32\,nm of the ion straggle. For the sample implanted with a fluence of 2$\times$10$^{15}$\,cm$^{-2}$, a Rutherford Back Scattering (RBS) profile is measured and compared to the simulation, Fig.\,\ref{fig1}\,(c); its maximum is found at 50\,nm from the surface with the straggle of 70\,nm. Thus at higher fluences, migration of the ions takes place during the annealing and, therefore, the distribution of the ions broadens. For lower fluences, the RBS is not measured as the ion density is below the detection limit.

To reduce the defect density, we apply two methods: increase of the substrate temperature during the implantation and variation of the post-implantation annealing parameters. To increase the substrate temperature during the implantation, a special sample-stage is designed with a heating element and a temperature detector with a temperature limit of 600\,K, which is close to the Debye temperature of Y$_2$SiO$_5$, $T_D=580$\,K \cite{sun2009}.
Annealing procedures are varied from the rapid thermal annealing (RTA) in N$_2$ atmosphere to the long thermal annealing in Argon and air. Annealing for several hours in Argon atmosphere or in air is found to be the most effective. An example of four implanted patterns is given in Fig.\,\ref{fig1}\,(d). Other samples are implanted with larger patterns of 1\,mm$\times$1\,mm in size. 

\textbf{Measurement technique}

Samples are characterized with 656\,nm CW laser to achieve the up-conversion excitation at room temperature and at 4\,K in a confocal setup \cite{siyushev2014}. Erbium luminescence is measured in reflection and directed to a grating spectrometer. The detection range is from 490\,nm to 650\,nm, see Fig.\,\ref{fig1}\,(e). For 4\,K measurement, sample is placed on a cold finger in a cryostat with access to the sample through a window. 
To measure relaxation times, samples are excited with pulses of 50\,$\mu$s and the following decrease of the intensity is recorded, \ref{app_B} and Fig.\,\ref{app_fig1}.

\section{Resonance energy transfer}
At low concentrations ions, the intensity of the luminescence is linearly proportional to the number of the ions. When the concentration increases, the non-radiative energy exchange between the ions becomes dominant, and the total emitted energy distributes between the radiative and the non-radiative processes \cite{daniels2003,andrews2009}. The last one is described by non-radiative resonant energy transfer (RET), which happens either due to a perfect match between the absorption and the emission energies of the neighbouring ions or with assistance of the ion-phonon interactions, which compensate mismatches of the ions energies. This effect is also known as a saturation and quenching of the luminescence intensity with the increase of concentration \cite{peak1983} and is described by the F\"{o}rster law \cite{forster1948}.

From the RET theory and F\"{o}rster law, the non-radiative energy transfer rate, W$_{NR}$, is due to the electric-dipole-electric-dipole (ED-ED) interactions and is proportional to $\textbf{R}^{-6}$ \cite{peak1983}:
\begin{equation}
	W_{NR} = C \frac{R_F^6}{\textbf{R}^6},
	\label{eq:1}
\end{equation}
where $\textbf{R}$ is the distance between the donor and acceptor ions, $C$ is the concentration of the acceptors, $R_F$ is the F\"{o}rster radius, at which energy on the radiative and non-radiative exchange is distributed equally. As we characterize optical transitions of Erbium, which are electric-dipole allowed \cite{weber1967,Li1992,dodson2012}, we consider energy transfer only due to the ED-ED interaction and neglect the higher order interactions and the interactions involving magnetic dipoles (MD). 
To apply the RET to our experimental data, we need to integrate the energy transfer probability for one donor-acceptor pair over the possible neighbours in the volume \cite{wolber1979,kim2008,lunz2011}:
\begin{equation}
	\Gamma= C \int{\int{\int{\frac{R_F^6}{\textbf{R}^6}}dr_x}dr_y}dr_z,
	\label{eq:2}
\end{equation}
with $\textbf{R}^2=r_x^2+r_y^2+r_z^2$. The ion-concentration varies over the implanted layer, Fig.\,\ref{fig1}\,(c), thus the interatomic distance $\textbf{R}$ varies as well. To simplify, we introduce a mean interatomic distance, $R_m$, which is derived from the implanted fluence $F$ and the ion straggle $d$: $R_m=\sqrt[3]{d/F}$, and an average concentration $C_{av}=F/d$, which is the concentration at half-maximum of the ion-distribution, see Fig.\,\ref{fig1}\,(c). This average concentration is twice smaller than concentration at the maximum of the ion-distribution peak. 
Thus, we obtain the following expression for the RET rate:
\begin{equation}
	\Gamma=C_{av}\,R_F^6 \int^\infty_{R_m}{\frac{1}{\textbf{R}^6}}d^3r= \frac{4\pi}{3}R_F^6\frac{F^2}{d^2}.
	\label{eq:3}
\end{equation}
The fit-function for the luminescence intensity on the fluence in this case has the following form:
\begin{equation}
	I=I_0\,S\,F\frac{1}{1+\Gamma}=I_0\,S\,F\frac{1}{1+\frac{4\pi}{3}R_F^6\frac{F^2}{d^2}},
	\label{eq:4}
\end{equation}
where $S$ is area of the measured spot and $I_0$ is intensity per ion.

At low ion-concentrations, RET is absent or negligibly small, and the luminescence intensity is proportional to the number of ions, and therefore to the fluence: $I=I_0\,S\,F$. When the ion density increases, the impact of RET on the luminescence yield becomes significant, and a part of the energy goes into the non-radiative energy exchange between the ions. In the result, the luminescence intensity per ion decreases with the number of ions as a function of their concentration: $I_0^*=I_0/(1+\Gamma)$.
This equations are further used to explain the experimental dependence of luminescence on fluence.

\section{Experimental results}

\textbf{Shapes of the spectra}

\begin{figure*}[t]
	\includegraphics[width=\textwidth]{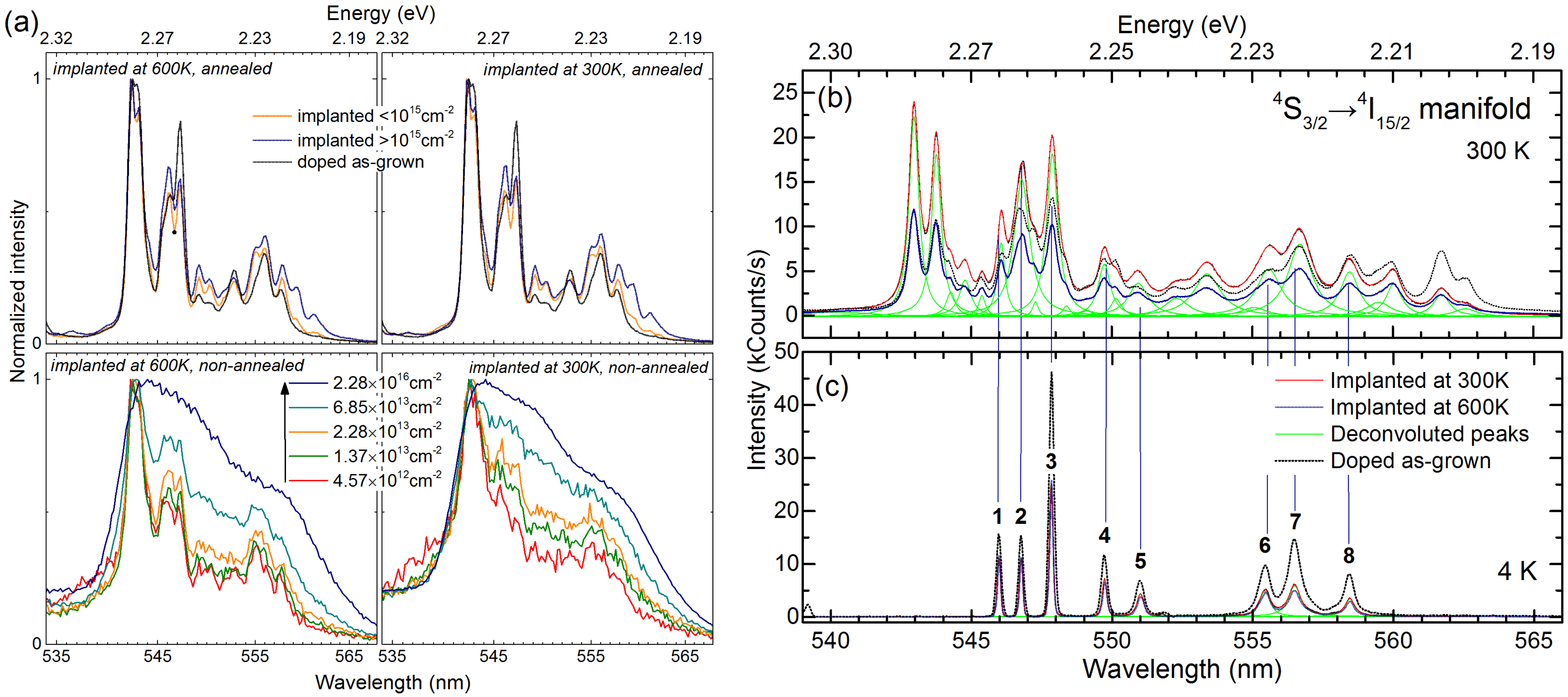}
	\caption{\textbf{(a)} Luminescence spectra of the $^4$S$_{3/2}\rightarrow^4$I$_{15/2}$ manifold for annealed and non-annealed samples implanted at temperatures of T\,=\,300\,K and 600\,K temperatures, indicated in each graph. All spectra are normalized to their maximal value. For the annealed samples, spectra are similar for all fluences.
		\textbf{(b-c)} Deconvolution of $^4$S$_{3/2}\rightarrow^4$I$_{15/2}$ manifold spectra of doped as-grown and implanted samples, which are measured at \textbf{(b)} 300\,K and \textbf{(c)} 4\,K. For the spectrum at 4\,K \textbf{(c)}, the independent lines are assigned to the transitions to specific sub-levels of the ground multiplet.}
	\label{fig2}
\end{figure*}
Luminescence study of the annealed and non-annealed samples is presented in Fig.\,\ref{fig2}.
The spectral shape of the Erbium optical transitions contains information about the surrounding crystal field and, consequently, about the ion activation in identical lattice sites. This information is given by the positions and linewidths of the spectral lines and overall shape of the spectra. Variety of local crystal fields from site to site results in a broad spectrum. This can be seen for the non-annealed samples in Fig.\,\ref{fig2}\,(a): For all non-annealed samples with the fluences above $10^{14}$\,cm$^{-2}$, spectra take a broad shape and specific sub-level transitions are not distinguishable. 
The implantation temperature is found essential at low fluences, when noticeable ion-activation without annealing is favoured. This effect can be seen in the normalized spectra in Fig.\,\ref{fig2}\,(a) for the fluences below $10^{14}$\,cm$^{-2}$ as more distinguished shapes of the emission lines of the non-annealed sample. For the room-temperature implanted samples, spectra are similarly broad at all fluences.

Annealing time and atmosphere have crucial effect on the stabilization of the ions in the lattice positions. We found that long annealing in air leads to an optical response, which is the most close to the doped as-grown sample. Comparison of the annealing procedures can be found in \ref{app_A}. Further, we discuss the results for the samples annealed in air.

For the air-annealed samples, the spectral shapes are independent of the fluence \footnote{This fact is correct for each annealing method.} and are similar for both implantation temperatures, Fig.\,\ref{fig2}\,(a). At the fluences above $10^{15}$\,cm$^{-2}$, several new emission lines appear while a few other are enhanced in intensity. This occurs as a result of the high density of ions and probable formation of clusters. Spectra of the implanted samples reveal differences from the doped as-grown crystal, which are similar over all fluences and implantation temperatures. This proves the reproducibility and stability of the implantation-annealing processes, which nevertheless lead to minor differences in local crystal symmetry as compared to the doped as-grown crystal.

\textbf{Spectral lines}

More detailed analysis of the implanted ions is made for the  manifold $^4$S$_{3/2}\rightarrow^4$I$_{15/2}$, as shown in Fig.\,\ref{fig2}\,(b,c). At room temperature, positions of the deconvoluted peaks are similar for all samples and are most closely described by a Lorentzian function. The minimal linewidth of the spectral lines at room temperature is found to be 1.32\,meV (319.0\,GHz) for the implanted samples and 1.29\,meV (311.7\,GHz) for the doped as-grown. The maximum linewidth at room temperature is 4.38\,meV (1054.5\,GHz) for the both implanted and doped as grown samples.

\begin{figure*}[ht]
	\includegraphics[width=\textwidth]{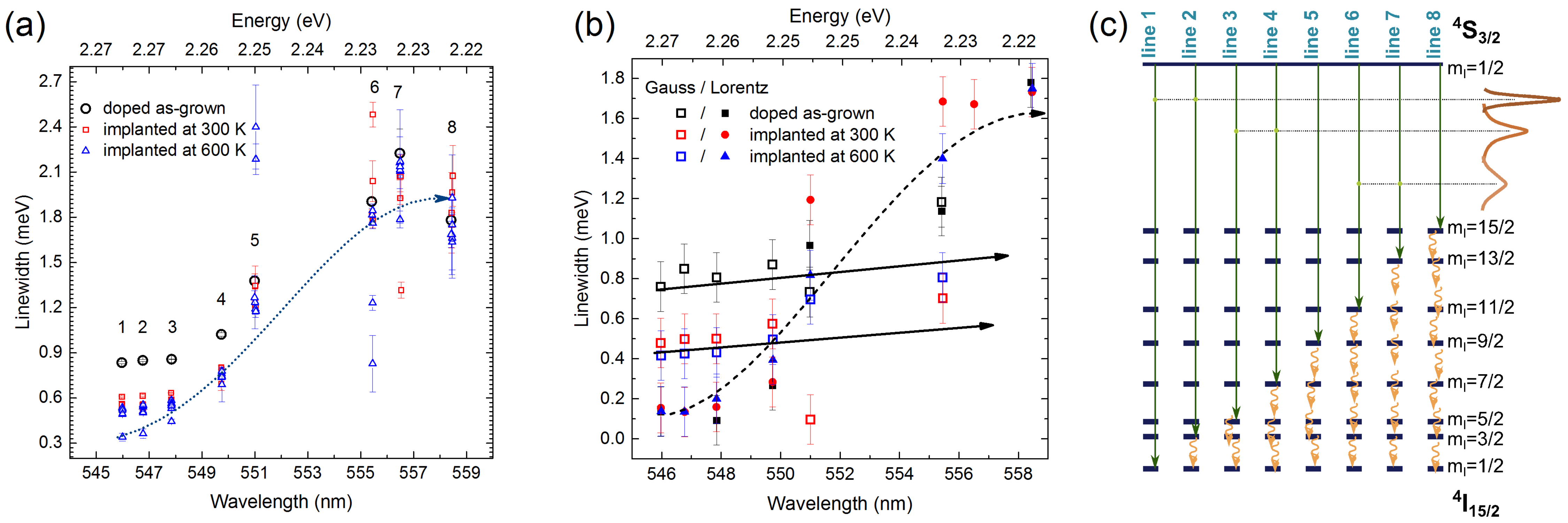}
	\caption{\textbf{(a)} Linewidth values for the eight transition lines of $^4$S$_{3/2}\rightarrow^4$I$_{15/2}$ manifold observed at 4\,K. thedotted line describes increase of the linewidth value with the wavelength. \textbf{(b)} Lorentzian and Gaussian components of the spectral lines of the $^4$S$_{3/2}\rightarrow^4$I$_{15/2}$ manifold. The Gaussian component remains relatively constant (solid line), while the Lorentzian component rises with the wavelength (dashed line). The increase of the Lorentzian component and thus the broadening of the spectral lines is suggested due to electron relaxation from the higher sublevels to the ground-state sublevel of the ground multiplet $^4$I$_{15/2}$, \textbf{(c)}.}
	\label{fig3}
\end{figure*}

At low temperatures (Fig.\,\ref{fig2}\,(c)), only eight transition lines remain. The lines labelled 1 and 2 exhibit purely inhomogeneous line shape and linewidths of 0.56\,meV (134.8\,GHz) for the implanted samples and of 0.83\,meV (201.3\,GHz) for the doped as-grown. However at longer wavelengths, lines become homogeneously broadened, and the linewidth increases to 1.53\,meV (369.7\,GHz) for the implanted samples and 1.64\,meV (396.3\,GHz) for the doped as-grown, as can be seen for the lines 5-8 in Fig.\,\ref{fig2}. 
All eight transition lines are associated with the transitions from the lowest level of the $^4$S$_{3/2}$ multiplet to the eight sublevels of the ground multiplet $^4$I$_{15/2}$ as shown in Fig.\,\ref{fig3}\,(c). They correspond to the Site\,2 of Erbium in YSO noted by B\"{o}ttger et al.\,\cite{bottger2006}. 

The variation of the linewidth and the line-shape with the wavelength at low temperature is demonstrated in Fig.\,\ref{fig3}\,(a). The very first transition line has a purely inhomogeneous, Gaussian shape, while for the other lines, a homogeneous, Lorentzian, component becomes more relevant and the linewidth increases. 

In order to separate the homogeneous and inhomogeneous broadening components and to see how their values vary with the wavelength, the low-temperature transition lines are fit with a Voigt fit-function, Fig.\,\ref{fig2}\,(c). The derived components are plotted in Fig.\,\ref{fig3}\,(b). The inhomogeneous component remains similar for all eight lines (as demonstrated with the solid lines in Fig.\,\ref{fig3}\,(b)). The homogeneous component appears to be significant in the third transition line and increases until it is completely dominant in the last three lines. 

Appearance of a large homogeneous broadening is explained with the transitions within the ground-state multiplet, as shown in Fig.\,\ref{fig3}\,(c). After a radiative transition to one of the sub-levels of the ground $^4$I$_{15/2}$ multiplet, an electron relaxes further to the lowest sub-level. Each transition line is assigned to a particular transition from the $^4$S$_{3/2}$ multiplet to the ground state sub-level as depicted in Fig.\,\ref{fig3}\,(c), and the homogeneous broadening part is found to be proportional to the energy separation between the higher-energy sub-levels of the ground multiplet and its lowest sub-level.



For the direct transition from $^4$S$_{3/2}$ to the lowest sub-level of the ground multiplet at low temperature, when phonon modes are ``frozen'', the natural linewidth (40\,kHz or 0.16$\times10^{-6}$\,meV) is dominated by an inhomogeneous broadening due to the variations in the crystal-field strains experienced by each ion, differences in the transition energies for different isotopes and/or impact from the hyperfine structure. 
Thus, in case of a larger inhomogeneous linewidth, ions are experiencing more differences in the surrounding crystal fields.
The first and the second transition lines in Fig.\,\ref{fig2}\,(c) have purely Gaussian shape. These lines are assigned to the transitions between the lowest state of the $^4$S$_{3/2}$ manifold and two lowest states of the $^4$I$_{15/2}$ manifold, Fig.\,\ref{fig3}\,(c). 

As mentioned earlier, the implanted samples show narrower linewidth as the doped as grown sample: 0.56meV versus 0.83meV, respectively. The width of the spectral line has contribution from hyperfine structure, isotope shifts and crystal field. Implanted Erbium has $\approx90\%$ of $^{170}$Er-isotope which can be the reason for a smaller linewidth. However, the inhomogeneous broadening of a transition line due to the presence of different isotopes is taken from Haynes et al. \cite{haynes1965, macfarlane1992} and equals to 0.02\,meV. The broadening due to the hyperfine structure of the Er$^{167}$-isotope is expected to be in the same order as the estimated isotopic broadening.
Therefore contributions from the hyperfine structure and isotopic broadening are neglected. The difference between the emission linewidths of the implanted and doped as-grown samples can be explained by formation of a more homogeneous local crystal field.

The destruction of the substrate by the implantation and recovery of it by the annealing creates a more homogeneous crystal structure and thus results in a smaller number of differences in the crystal-fields. This more homogeneous crystal structure is not due to lower number of defects in the lattice, but due to a more homogeneous defect distribution. 

However at room temperature, the linewidth in the doped as-grown sample is smaller as compared to the implanted one. The dependence of the linewidth on the temperature is commonly described by the McCumber-Sturge relation \cite{mccumber1963}:
\begin{equation}
	\Gamma(T)=\Gamma_0+\alpha\left(\frac{T}{T_D}\right)^7\int_0^{T_D/T}\frac{x^6\times e^x}{(e^x-1)^2}dx,
\end{equation}
where $T_D$ is the Debye temperature of the material, $\Gamma_0$ is the linewidth at 0\,K temperature, $T$ is the actual temperature and $\alpha$ is the coefficient proportional to the strength of the electron-phonon interaction \cite{mccumber1963,vanvleck1940,liu2015}. Thus, the influence of the phonons on the spectral lines appears in the value of $\alpha$. 
From the known linewidth values, we derive values of $\alpha$ for our samples. For the doped as-grown sample, $\alpha=93$\,cm$^{-1}$, whereas for the implanted sample it is equal to 155\,cm$^{-1}$. This difference in the coefficients indicates that the ion-phonon coupling is much stronger in the implanted samples. The stronger ion-phonon coupling can be interpreted as a confirmation of a specific homogeneous defect-entourage, which leads to the narrower spectral lines.

\textbf{Implantation yield}

Not every implanted ion can settle in a desired lattice site. Ratio between the implanted fluence and the final amount of the activated ions is described by the implantation yield. It can be derived from the number of luminescent ions normalized to the total number of the implanted ions. We estimate the yield by comparing the luminescence of the doped as-grown sample to the implanted sample and by taking the ratio of the concentrations corresponding to equal intensities, as described in \cite{kukharchyk2014}. The measured volume is assumed to be equal to 300\,nm$\times$300\,nm $\times$1\,$\mu$m, and thus the implantation yield is found to be 20\% for the room-temperature implantation and 22\% for the hot implantation.

\textbf{Energy transfer between the implanted ions}

\begin{figure*}[ht]
	\includegraphics[width=\textwidth]{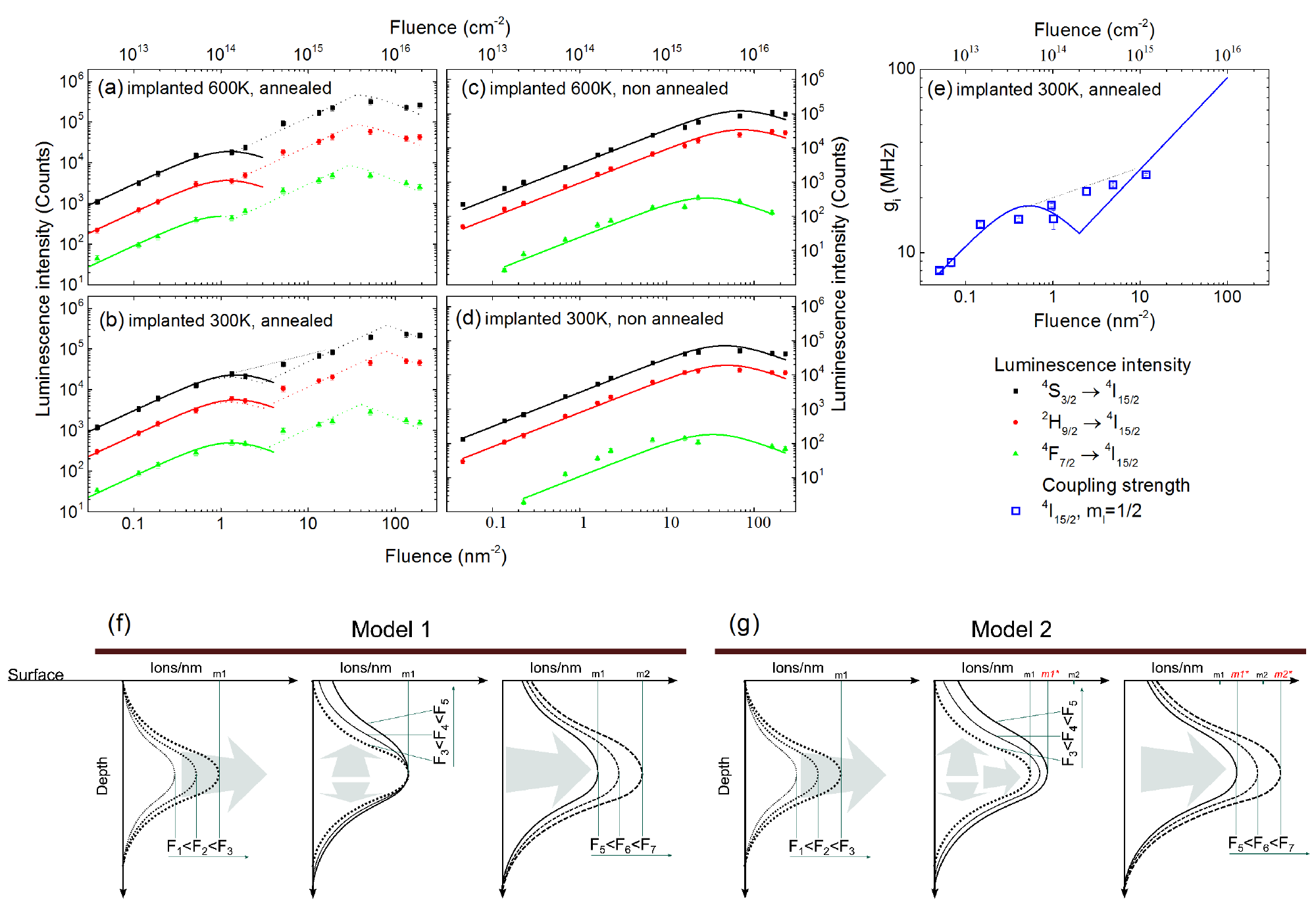}
	\caption{The luminescence intensity as a function of the implanted fluence is shown for the implanted at 600\,K, \textbf{(a)} annealed and \textbf{(b)} non-annealed samples, also for implanted at 300\,K, \textbf{(c)} annealed and \textbf{(d)} non-annealed samples. Fit curves are shown with the solid and dotted lines for three multiplets: $^4$F$_{7/2}\rightarrow^4$I$_{15/2}$ (green), $^2$H$_{11/2}\rightarrow^4$I$_{15/2}$ (red) and $^4$S$_{3/2}\rightarrow^4$I$_{15/2}$ (black). Solid lines relate to the fit with a single function, Eq.\,(\ref{eq:4}). Dotted lines describe the fit with Model 1. 
		\textbf{(e)} Application of RET model to the collective coupling strength, reported earlier by Probst et.al. \cite{probst2014}.
		\textbf{(f)} Schematic illustration of Model 1: after exceeding the critical fluence F$_3$, ion-straggle increases while the concentration in the maximum (m1) remains the same.
		\textbf{(g)} Schematic illustration of Model 2: after the critical fluence F$_3$ is exceeded, ion-straggle increases as well as the concentration in the maximum (m1$\rightarrow$m1$^*$) of the ion distribution.
	}
	\label{fig4}
\end{figure*}
For the previous calculation of the implantation yield, we plotted the integrated luminescence intensity as a function of fluence, Fig.\,\ref{fig4}. The intensity is integrated over three manifolds separately: $^4$F$_{7/2}\rightarrow^4$I$_{15/2}$, $^2$H$_{11/2}\rightarrow^4$I$_{15/2}$ and $^4$S$_{3/2}\rightarrow^4$I$_{15/2}$. For each manifold, the dependence is analysed according to Eq.\,(\ref{eq:4}).

For the annealed samples, two saturation steps are observed, as shown in Fig.\,\ref{fig4}\,(a,b). the first saturation step appears at a fluence around $10^{14}$\,cm$^{-2}$, after which the intensity continues to increase with the fluence until it is completely saturated at fluences of $10^{16}$\,cm$^{-2}$.
To explain these two steps, we consider the limitation of the Erbium concentration in YSO for growth, which is approximately 10\% of the total number of Yttrium ions \cite{isshiki2003} and equals $C_{th}=2.4\times10^{20}$\,cm$^{-3}$. This concentration at the maximum of the ion-distribution corresponds to a fluence of $3.84\times10^{14}$\,cm$^{-2}$. If the 10\%-limit is exceeded, creation of Er-Si-O formations is observed \cite{isshiki2003}.

With respect to this, we explain the two saturation steps as following: At fluences of $10^{14}$\,cm$^{-2}$, the local volume concentration reaches the growth limit of $C_{th}$. Ions are embedded into the lattice so close to each other that RET takes place, and a saturation step is observed. However during the annealing, the crystal structure tends to stay as Y$_2$SiO$_5$ crystal without cracks and clusters. This results in migration of the implanted ions to the closest vacant positions created by implantation, which are dislocations in the damaged layer (distribution shown in Fig.\,\ref{fig1}\,(c)). As a result, the highest ion concentration remains relatively same while the ion-straggle $d$ increases, and therefore the total luminescence intensity increases with luminescent yield per ion (and concentration) remaining relatively constant, Fig.\,\ref{fig4}\,(f). The diffusion of the ions is limited by the thickness of the available damaged layer, which equals to the sum of the depth of ion-maximum, $h_{max}=50$\,nm, and half of the ion-straggle, $d/2=16$\,nm, see Fig.\,\ref{fig1}\,(c). The result of Ratherford back scattering for a sample implanted with a fluence of $2\times10^{15}$\,cm$^{-2}$ is presented in Fig.\,(\ref{fig1}\,(c)) and is the experimental proof of the diffusive behaviour.

When the avaliable layer appears to be completely filled with threshold concentration of $C_{th}$, the reached straggle $d_{max}$ becomes constant, and the concentration begins to increase again. With this rise of concentration, luminescence quenching is observed again, yielding into a continuation of the previous slope. We name this behaviour as Model 1, which is schematically demonstrated in Fig.\,\ref{fig4}\,(f).

Experimental data for the annealed samples are fitted with a function based on Eq.\,(\ref{eq:4}), which is adapted for the Model 1: for the fluences below the threshold fluence $F_{th}\approx10^{14}$\,cm$^{-2}$, the concentration changes proportionally with the fluence; above the $F_{th}$, the concentration remains constant until the volume limit is reached, after which the concentration increases proportionally with the fluence again. Details on the fit functions are given in \ref{app_C}. 
However, above the threshold fluence $F_{th}$, the concentration may still be slowly increasing, simultaneously with the ion-straggle, Model 2. In this case, the intensity depends on both the straggle and the fluence, and its slope changes. This appears to be the case for the room temperature implantation, Fig.\,\ref{fig4}\,(b), while hot implantation is well-described with Model 1, Fig.\,\ref{fig4}\,(a).

The solid line in Fig.\,\ref{fig4}\,(a,b) shows the fit until fluences of $2\times10^{14}$\,cm$^{-2}$ using Eq.\,(\ref{eq:4}). 
By fitting Model 1 to the full range of fluences, we obtain good agreement with the data, as shown by the dotted line.
For the both fits, the F\"{o}rster distance is derived from the fit to be equal to 2.31\,nm and 2.18\,nm for the hot and room-temperature implantations, respectively. 
These values are similar to the F\"{o}rster distances numerically evaluated in the literature for the transitions between $^4$I$_{13/2}$ and $^4$I$_{15/2}$, which are approximately 1.73\,nm \cite{tarelho1997,desousa2002}. 
The F\"{o}rster distances are expected to vary for different manifolds, however, they should stay in the same order of values.
The threshold fluence for the hot implantation is equal to $1.5\times10^{14}$\,cm$^{-2}$, and for the room-temperature implantations it is equal to $3\times10^{14}$\,cm$^{-2}$. Earlier luminescence saturation for the hot implantation is related to the higher activation rate of the ions at the crystallographic sites during the implantation and, thus, higher final concentration at equal fluences. 

For the non-annealed samples, we observe only one saturation step at fluences of 10$^{16}$\,cm$^{-2}$. The energy transfer in the annealed samples can be assisted by phonons. Therefore, it can occur for larger mismatches between the transition energies. In non-annealed areas, a large variation of the crystal-field strengths reduces matching between the emission and absorption energies of the neighbouring ions. Additionally, there are very few phonon energy-states present in the damaged crystal lattice, and, therefore, it requires a much higher fluence to achieve the luminescence saturation effect.

For the non-annealed samples, we obtain the F\"{o}rster distance of 0.54\,nm for the room temperature implantation, Fig.\,\ref{fig4}\,(d), and 0.61\,nm for the hot implantation, Fig.\,\ref{fig4}\,(c), with the ion straggle $d$ taken from SRIM as 32\,nm, Fig.\,\ref{fig1}\,(c). These values are already close to the smallest possible distances between the two ions in the lattice, which is approximately 0.5\,nm. At such distances, luminescence quenching may happen via an electron transfer (Dexter energy transfer) \cite{dexter1953}. The luminescence intensity per ion in the non-annealed samples is higher in the case of hot implantation, which is again due to the partial direct activation of the ions in the lattice.

Model 1 and Model 2 are also supported by the dynamics of the relaxation time value. The relaxation time shortens as the concentration of ions and the number of interactions between them increase \cite{Li1992}. At the lowest fluences, our measured relaxation times are equal to 4.2\,$\mu$s and reduce to 4.0\,$\mu$s at the fluences of $10^{14}$\,cm$^{-2}$. Their values remain equal to 4.0\,$\mu$s in entire diffusion range. Around the final saturation step, the relaxation times decrease and reach 3.2\,$\mu$s.

The effects of RET and the concentration non-linearity are also observed in circuit QED experiment. There, an array of superconducting resonators is magnetically coupled to the room-temperature implanted sample \cite{probst2014}, Fig.\,\ref{fig4}\,(e). The dependence of the collective coupling strength on the fluence is in agreement with Model 2. Accordingly, we obtain the modified fit-function for the collective coupling strength:
\begin{equation}
	g_{col}=g_0 \sqrt{S_{res} F \frac{1}{1+\frac{4 \pi}{3}R_{Fm}^6\frac{F^2}{d^2}}},
\end{equation}
where $g_{col}$ is the collective coupling strength, $g_0$ is the coupling strength per ion, $S_{res}$ is the area under the resonator and $R_{Fm}$ is the F\"{o}rster distance for the magnetic-dipole-magnetic-dipole interaction.
Due to the lack of data-points in the region from $10^{15}$\,cm$^{-2}$ to $10^{17}$\,cm$^{-2}$, we cannot properly fit the data by Model 1. 
Nevertheless, the solid line in Fig.\,\ref{fig4}\,(e) shows the closes correspondance to the data. Value of the F\"{o}rster distance in this case is found to be equal to 2.76\,nm, which is larger than 2.18\,nm for the optical transitions. Thus, the magnetic dipole interaction occurs at larger distances between the ions. Nontheless, it explains the deviation of the coupling strength dependence on the number of ions at rather low fluences \cite{probst2014}. 

\section{Conclusion}
We studied the ion implantation process in Y$_2$SiO$_5$ crystals in combination with thermal annealing. We found that the proposed method produces stable and repetitive results for the ion activation in the crystals. The implantation yield for Erbium ions is estimated to be equal to 20\%$\pm$5\%. The luminescence studies suggest that the implanted ions experience more homogeneous crystal field as compared to the doped as-grown samples. However, the such homogeneous crystal field is accompanied by a larger number of defects, which are also stronger coupled to the Erbium ions. The luminescence intensity and collective coupling strength of implanted er ions reveals a non-linear dependence above the fluence of 10$^{14}$\,cm$^{-2}$. 
Employing the resonance energy transfer theory, we developed two models describing the observed non-linear behaviours.
The presented research paves the way towards a feasible technological process for fabrication of quantum-memory elements in hybrid circuits.

\appendix
\section{Annealing procedures}
\label{app_A}
\begin{figure}[t]
	\includegraphics[width=\linewidth]{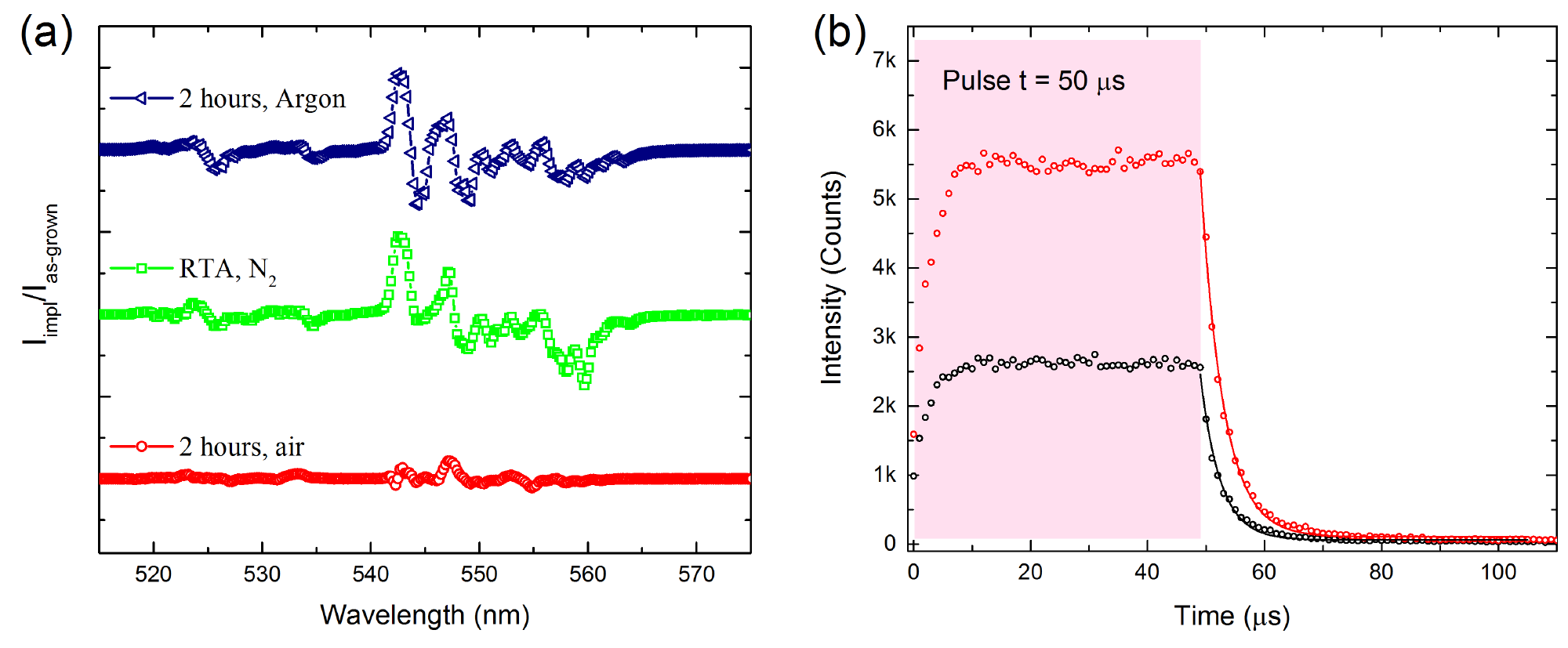}
	\caption{\textbf{(a)} Spectra of the samples annealed in different atmospheres. Spectrum of each sample is normalized by the spectrum of doped as-grown sample, $I_{imp}/I_{as-grown}$. \textbf{(b)} An example of the relaxation-time measurement.}
	\label{app_fig1}
\end{figure}

Several annealing procedures were applied for the post-implantation treatment. Rapid thermal annealing (RTA) was performed in N$_2$ atmosphere with a duration of 8 minutes. Longer annealing lasted for 2 hours in two different atmospheres: air and Argon. Air was selected as oxygen-enriched atmosphere.
Efficiency of the procedure is concluded from the similarity between optical spectra of the implanted and doped as-grown samples. Spectra of the implanted samples are normalized by the spectrum of the doped as-grown sample, as shown in Fig.\,\ref{app_fig1}\,(a). For the air-annealed samples (independently on the implantation temperature), normalized spectrum results in almost flat line. For the other samples, a lot of deviations are observed. Thus, the air-implanted sample has the local symmetries most closely matching to the structure of the doped as-grown sample.

\section{Relaxation time}
Relaxation times as derives from pulsed measurements, when samples are exited with pulses of 50\,$\mu$s, see Fig.\,\ref{app_fig1}\,(b). Following decay of intensity is fit with an exponential function, and thus the relaxation time values are derived.

\section{Resonance energy transfer.}
\label{app_B}
For the RET due to electric-dipole-electric-dipole interaction, the quantum amplitude is given by:
\begin{equation}
	M^{e-e}_{fi}\bar{M}^{e-e}_{fi}=|\mu^A|^2|\mu^B|^2 \frac{2}{(4 \pi\epsilon_0 R^3)^2}(3+k^2R^2+k^4R^4)
\end{equation}
where $\mu$ is the electric dipole operator and $R$ is the interatomic distance, $k$ is the wave vector, and A and B correspond to the donor and acceptor atoms. For the transitions at 560\,nm, the wave vector $k=1/560$\,nm$^{-1}$, and $k R << 1$ at the interatomic distances below 20\,nm, thus the $k^2R^2$ and $k^4R^4$ terms can be neglected. So, we have only the $1/R^6$ dependence:
\begin{equation}
	M^{e-e}_{fi}\bar{M}^{e-e}_{fi}=|\mu^A|^2|\mu^B|^2 \frac{6}{(4 \pi\epsilon_0 R^3)^2}.
\end{equation}
In a similar way, the quantum amplitude of the RET due to the magnetic-dipole-magnetic-dipole interaction is as follows:
\begin{equation}
	M^{m-m}_{fi}\bar{M}^{m-m}_{fi}=|m^A|^2|m^B|^2 \frac{2}{c^2(4 \pi\epsilon_0 R^3)^2}(3+k^2R^2+k^4R^4),
\end{equation}
where $m$ is the magnetic dipole operator. This similarly can be reduced to the $R^{-6}$ dependence as above. 
For the electric-dipole-magnetic-dipole interaction, the main contribution comes from the term $k^2R^{-4}$, which is weaker than the $R^{-6}$ an thus can be neglected in our calculations.

The transfer rate function is derived from the $R^{-6}$ dependence normalized by the F\"{o}rster radius, which corresponds to the equal energy distribution between the radiative and the non-radiative energy transfer, thus the latter one equals
\begin{equation}
	\Gamma = \frac{R_F^6}{R^6},
\end{equation}
and the final dependence for a single pair of atoms has the following form:
\begin{equation}
	I_{rad} = I_{tot}\frac{1}{1+\Gamma}=I_{tot}\frac{1}{1+\frac{R_F^6}{R^6}}.
\end{equation}

\section{The fit function. }
\label{app_C}
The RET fit-function is used in the following form:
\begin{equation}
	\mathbf{y}=y_0 \frac{\pi 150^2 \mathbf{x}}{1+4\pi r_F^6 \mathbf{x}^2 / 3d_0^2},
\end{equation}
when the fluence $\mathbf{x}$ is smaller than the threshold fluence $\mathbf{x}<F_{th}$. Above the threshold fluence, ion straggle $d$ increases and also becomes a variable: 
\begin{equation}
	\mathbf{y}=y_0 \frac{\pi 150^2 \mathbf{x}}{1+4\pi r_F^6 \mathbf{x}^2 / 3\mathbf{d}^2}.
	\label{eqff1}
\end{equation}
In model 1, while the straggle increases, the concentration remains constant and the function reads:
\begin{equation}
	\mathbf{y}=y_0 \frac{\pi 150^2 \mathbf{x}}{1+4\pi r_F^6 F_{th}^2 / 3 d_0^2},
\end{equation}
so the radiative energy transfer is again proportional only to the ion fluence. In the model 2, both fluence and ion straggle keep changing, which is described by the Eq.\,(\ref{eqff1}).
Finally, when the possible volume is filled and the straggle does not change with fluence, the dependence retraces to its initial form:
\begin{equation}
	\mathbf{y}=y_0 \frac{\pi 150^2 \mathbf{x}}{1+4\pi r_F^6 \mathbf{x}^2 / 3d_{max}^2},
\end{equation}
where $d_{max}$ is the final maximum ion straggle.

\section*{Acknowledgement}
We would like to acknowledge Ronna Neumann for the fabrication of the LMIS and Georg Kortenbruck for the design of the sample-stage with the integrated heater.
We would like to acknowledge Benjamin Feldern valuable comments on the manuscript.
Additionally, S.Pal and A.D.Wieck would like to acknowledge IMPRS Surmat, MPIE Düsseldorf.
This work was financially supported by the BMBF program ``Quantum communications'' through the project QUIMP.

\bibliographystyle{apsrev4-1}
\bibliography{main}

\end{document}